\documentclass[9pt,amsmath,amssymb,noshowpacs,nofootinbib]{revtex4}
\usepackage{amsmath,amsfonts}
\usepackage{graphicx}
\usepackage{hyperref}

\begin{document}
\title{Fun from none: deformed symmetries and Fock space}
\author{Michele Arzano}
\email{marzano@perimeterinstitute.ca}
\affiliation{Perimeter Institute for Theoretical Physics\\
31 Caroline St. N, N2L 2Y5, Waterloo ON, Canada\\
and\\
Institute for Theoretical Physics,\\
Utrecht University, Leuvenlaan 4,\\
Utrecht 3584 TD, The Netherlands}

\begin{abstract}
\vskip .2cm
\begin{center}
{\bf Abstract}
\end{center}
We give a pedagogical introduction to the basics of deformations of relativistic symmetries and the Hilbert spaces of free quantum fields built as their representations.  We focus in particular on the example of a $\kappa$-deformed scalar quantum field for which the generators of spatial translations that label the field modes act according to a deformed Leibnitz rule.  We explore the richer structure of the $\kappa$-Fock space and point out possible physical consequences of the deformation.
\end{abstract}
\maketitle

\section{Introduction}
In ordinary applications of quantum field theory (QFT) to high energy physics non-interacting fields are rather trivial entities which describe particles for which no scattering can take place.  As soon as we leave our laboratory and let the quantum fields propagate on a non-trivial background or, alternatively, accelerate our experimental apparatus the free field comes alive and surprising phenomena of particle production take place.  Ultimately these phenomena can be explained in terms of the existence of inequivalent field quantizations: each observer will define a notion of energy according to her own time evolution and this in turn determines the vacuum state of her own Fock space \cite{Ashtekar:1975zn}.\\ 
A less known context in which an ordinary free field exhibits non-trivial features upon quantization is provided by the multi-particle sector of a {\it deformed} Fock space where a $n$-particle state is no longer a (anti)symmetrized $n$-tensor product of one-particle states (for (fermions) bosons).  Such deformed tensor products still provide a representation of the symmetric group but now the statistics becomes {\it momentum dependent} \cite{Arzano:2008bt} in a way which will become clear in the following.  In the present contribution we attempt to give a pedagogical introduction to such models and show how deformed Fock spaces naturally emerge when the phase space of the field carries a representation of certain quantum deformations of the (Hopf) algebra of relativistic symmetry generators.  These deformed algebras have been proposed as emerging from the ``no-gravity" limit of quantum gravity i.e. as to effectively represent the isometries of a ``quantum version" of Minkowski space and are usually introduced to preserve covariance properties of certain non-commutative space-times (see e.g. \cite{Majid:1994cy, Bais:2002ye, Oeckl:2000eg, KowalskiGlikman:2002jr, AmelinoCamelia:2003xp, Chaichian:2004za, Freidel:2005me}).\\
In the next Section we recall the basics of free field quantization emphasizing the main ingredients and mathematical structures underlying the construction of bosonic Fock spaces.  We then briefly comment on the simplest example of quantum deformation of relativistic symmetry generators, the ``twisted" Poincar\'e algebra and the states of the related quantum scalar field.  We then move our focus to $\kappa$-deformations of the Poincar\'e algebra and the quantization of a free scalar field.  We conclude with a discussion on the possible ``planckian" decoherence effects that the richer structure of the $\kappa$-Fock space can induce.

\section{Field quantization}
The states of a system described by a classical scalar field are represented by {\it points} in its phase space.  Usually the latter is given by the space $\Gamma_{\Sigma}$ of smooth initial data $\{\varphi,\pi\}$ of compact support on a given Cauchy surface $\Sigma$.  Observables are simply {\it functions} on $\Gamma_{\Sigma}$ the most important of them being the Hamiltonian which encodes the dynamics of the system.  If we want to describe the states of a composite system given, for example, by two subsystems $A$ and $B$ we simply take the {\it direct sum} of the individual phase spaces $\Gamma^{A\cup B}_{\Sigma}\equiv\Gamma^A_{\Sigma}\oplus\Gamma^B_{\Sigma}$.\\  For a quantum field things are radically different.  In this case the quantum states are {\it rays} in a complex Hilbert space $\mathcal{H}$ while observables are {\it self-adjoint operators} on $\mathcal{H}$.  At the level of composite systems the Hilbert space is now given by the {\it tensor product} of the subsystems Hilbert spaces $ \mathcal{H}^{A\cup B}\equiv \mathcal{H}^A\otimes\mathcal{H}^B$.\\  Field quantization is simply a recipe for obtaining the quantum description of the field system from its classical counterpart.  It turns out that such a transition is better understood in terms of the so-called {\it covariant phase space} formalism \cite{Ashtekar:1975zn, Crnkovic:1987tz}.  In this framework the phase space of our scalar field is taken to be the space of solutions of the Klein-Gordon equation $\mathcal{S}_{KG}$.  Indeed in Minkowski space (and, in general, for any globally hyperbolic space) each pair of initial data on a given space-like hyper-surface is uniquely associated to a solution of the equation of motion and this establishes an isomorphism between $\Gamma_{\Sigma}$ and $\mathcal{S}_{KG}$ as vector spaces.  The first step for obtaining a complex Hilbert space from $\mathcal{S}_{KG}$ is, as expected, to complexify the latter i.e. consider $\mathcal{S}^{\mathbf{C}}_{KG}\simeq\mathcal{S}_{KG}\otimes\mathbf{C}$ where $\mathbf{C}$ is the field of complex numbers.  In order to turn such complex vector space into a Hilbert space we need to introduce an inner product.  The vector space $\mathcal{S}_{KG}$ has a natural symplectic structure i.e. a (conserved) non-degenerate anti-symmetric bilinear form given by the following integral over a Cauchy hypersurface $\Sigma_t$ at fixed time $t$ 
\begin{equation}
\omega(\phi_1,\phi_2)= -\int_{\Sigma_t}(\phi_2\dot{\phi}_1-\phi_1\dot{\phi}_2)\, d^3\vec{x}\, ,
\end{equation}
which is nothing but the Wronskian for solutions of the Klein-Gordon equation.  The hermitian inner product on $\mathcal{S}^{\mathbf{C}}_{KG}$ will be given by
\begin{equation}\label{qftinnprod}
<\phi_1,\phi_2>\equiv -i\omega(\bar{\phi_1}, \phi_2)\, .
\end{equation}
where the bar indicates complex conjugation.
To complete the Hilbert space construction we need to restrict to a subspace of $\mathcal{S}^{\mathbf{C}}_{KG}$ on which the inner product above is positive definite, this will turn out to be the ``positive energy" subspace $\mathcal{S}^{\mathbf{C}+}_{KG}$.\\ %spanned by the solutions whose Fourier transform has support on the positive mass-shell  
The ``one-particle" Hilbert space of the theory $\mathcal{H}$ will be given by the Cauchy completion of $\mathcal{S}^{\mathbf{C}+}_{KG}$ with respect to the inner product (\ref{qftinnprod}).  The last step will be to construct, using $\mathcal{H}$ as the building block, the full ``multi-particle" sector of the theory.  As we mentioned above composite quantum systems are represented by tensor products of the Hilbert spaces of their components.  When we deal we systems of identical particles their quantum indistinguishability requires that we represent their states with combinations of tensor products of their Hilbert spaces which are symmetric under exchange of their labels.  In other words such states must carry a representation of the symmetric group.  For our quantum scalar field ``$n$-particle states" will be given by ``symmetrized" $n$-tensor products of the one-particle space $\mathcal{H}$ i.e. their Hilbert space can be written as
\begin{equation}
S_n\mathcal{H}^n=\frac{1}{n!}\sum_{\sigma\in P_n}\sigma(\mathcal{H}^{\otimes n})\, .
\end{equation}
where $\sigma $ a permutation in the permutation group of $n$-elements $P_n$ and $\mathcal{H}^{\otimes n}$ is the $n$-fold tensor product of $n$-copies of $\mathcal{H}$.  Given a ``one-particle" observable $\mathcal{O}$, i.e. a self adjoint operator on $\mathcal{H}$, its action on multiparticle states is given by the {\it second quantized} operator \cite{Geroch:1985ci}
\begin{equation}
d\Gamma(\mathcal{O})\equiv 1+ \mathcal{O} + (\mathcal{O}\otimes 1+1\otimes \mathcal{O})
 + (\mathcal{O}\otimes 1\otimes 1 + 1\otimes \mathcal{O}\otimes 1+ \ 1\otimes 1\otimes \mathcal{O})+...
\end{equation}
Such expression is simply telling us that the operator $\mathcal{O}$ acts on multiparticle states as a derivative i.e. following the Leibnitz rule.  However this quite obvious expression has an important mathematical significance.  Indeed the expression above can be re-written in terms of the {\it coproduct} $\Delta \mathcal{O}=\mathcal{O}\otimes 1 + 1\otimes \mathcal{O}$ of the operator $\mathcal{O}$
\begin{equation}
d\Gamma(\mathcal{O})\equiv 1+ \mathcal{O} + \Delta \mathcal{O} + \Delta_2 \mathcal{O}  + ... + \Delta_n \mathcal{O} + ...
\end{equation}
where $\Delta_n \mathcal{O}=(\Delta \otimes 1)\circ \Delta_{n-1}$, $\Delta_1\equiv \Delta$ with $n\geq 2$.
The coproduct introduced above is telling us that upon ``second quantization" the (associative) algebra of observables  acquires an additional structure which eventually turns it into a Hopf algebra.  In an effort to keep our presentation more straightforward we will not comment further on the role of Hopf algebra at this point (the interested reader can refer to \cite{Chari:1994pz} for an introduction to the subject).  The main point that should be kept in mind is that the coproduct encodes information on how observables act on a multi-component {\it quantum} system and, in our specific case, on the additivity properties of observables for the multiparticle sector of a quantum field Hilbert space.

\section{Beyond Leibnitz}
What has been said so far seems to have little to do with deformations of space-time symmetries.  This is not the case however.  Symmetry generators are just a special kind of observables.  Indeed, as we showed in the previous Section, the one-particle Hilbert space $\mathcal{H}$ is constructed from the space of solutions of the Klein-Gordon equation $\mathcal{S}$ and this in turn implies that $\mathcal{H}$ carries a unitary irreducible representation of the Poincar\'e algebra $\mathcal{P}$.  In particular we have a natural action of the generators of $\mathcal{P}$ as {\it one-particle operators}.  Moreover one-particle states can be completely characterized by a commuting set of such operators, for example in the standard plane-wave representation the eigenvalues of spatial translation generators ${\bf P}$ will label the orthonormal set of kets $|{\bf p}\rangle$. The coproduct $\Delta {\bf P}$ we introduced above extends the action of $\mathcal{P}$ to multiparticle states.  In usual QFT the information carried by such object is simply a restatement of the Leibnitz rule for the action of symmetry generators (and general observables) on tensor product states.  {\it Symmetry deformation provides a generalization of the additivity of symmetry generators encoded in the Leibnitz rule}\footnote{In this sense a deformed algebra of relativistic symmetries evades the Coleman-Mandula theorem whose proof assumes additivity.}.  A whole class of deformed symmetries can be obtained by introducing a ``modulation" of the coproduct weighted by a certain  {\it deformation scale} $q$
\begin{equation}
\Delta_{q}=\mathcal{F}_{q}\Delta\mathcal{F}^{-1}_{q}
\end{equation}
in such a way that given a symmetry generator $G$:
\begin{equation}
\Delta_{q} G(|\alpha\rangle\otimes |\beta\rangle)\neq\Delta_{q} G(|\beta\rangle\otimes |\alpha\rangle)
\end{equation}
i.e. {\it realize a non-symmetric Leibnitz rule for $G$ acting on multiparticle states}.  From our point of view space-time symmetry deformation is a {\it purely quantum (relativistic) phenomenon}:  we are not deforming relativistic kinematics {\it tout court}, we are simply changing the way classical relativistic symmetries are implemented in the multiparticle sector of (free) QFT.\\  Before discussing in detail some explicit examples of symmetry deformation we briefly review the basics of Fock space formalism.  We start with a plane wave basis $\{e_{\bf k}\}$ for the one-particle Hilbert space $\mathcal{H}$.  In terms of such vectors a generic one-particle state $\xi\in\mathcal{H}$ can be written as
\begin{equation}
 \xi=\sum_{\bf k}\xi({\bf k})|{\bf k}\rangle \equiv \sum_{\bf k}\xi({\bf k}) e_{\bf k}\, . 
\end{equation}
Accordingly a $n$-particle state will read
\begin{equation}
\xi_n=\frac{1}{\sqrt{n!}}\sum_{\bf k_1...k_n}\xi_n({\bf k_1...k_n})|{\bf k_1}...{\bf k_n}\rangle 
\end{equation}
where $|{\bf k_1}...{\bf k_n}\rangle$ indicates a symmetrized tensor product of basis elements $|{\bf k_1}...{\bf k_n}\rangle=\frac{1}{\sqrt{n!}}\sum_{\sigma\in P_n}e_{\bf k_1}\otimes...\otimes e_{\bf k_n}$.  We write a generic Fock space element as $\Psi = \{\xi_0, \xi_1({\bf k_1}),...\xi_n({\bf k_1},...,{\bf k_n}),... \}$.  We can now introduce annihilation and  creation operators whose action is defined respectively by 
$a_{\bf k}$ and $a^{\dagger}_{\bf k}$:
\begin{multline}
a_{\bf k}\, \{\xi_0, \xi_1({\bf k_1}),...\xi_n({\bf k_1},...,{\bf k_n}),... \}=\\
= \{\sum_{\bf k'} \delta_{\bf k k'} \xi_1({\bf k'}), \sqrt{2} \sum_{\bf k'} \delta_{\bf k k'} \xi_2({\bf k'},{\bf k_1})... \sqrt{n+1} \sum_{\bf k'} \delta_{\bf k k'} \xi_{n+1}({\bf k'},{\bf k_1},...,{\bf k_{n+1}}),...\}
\end{multline}
and
\begin{multline}
a^{\dagger}_{\bf k}\,\,\{\xi_0, \xi_1({\bf k_1}),...\xi_n({\bf k_1},...,{\bf k_n}),... \}\\
 = \{0,  \xi_0\, \delta_{\bf k k_1}, \sqrt{2}\,\, \mathrm{sym}(\delta_{\bf k k_1} \xi_1({\bf k_2})),... \sqrt{n}\,\, \mathrm{sym}(\delta_{\bf k k_1} \xi_{n-1}({\bf k_2},...,{\bf k_n})),... \}\, ,
\end{multline}
notice the symmetrization appearing in the last definition.  From these two definitions it is easily seen that the operators $a_{\bf k}$ and $a^{\dagger}_{\bf k}$ obey the canonical commutation relations
\begin{equation}
[a_{\bf k},a_{\bf k'}]=[a^{\dagger}_{\bf k},a^{\dagger}_{\bf k'}]= 0\, ,\,\,\,\,
[a_{\bf k},a^{\dagger}_{\bf k'}]= \delta_{\bf k k'}.
\end{equation}
Having set the stage we proceed in the next subsection with a quick overview of a rather popular type of deformation, the twisted Poincar\'e algebra.

\subsection{Twisted Fock space}
The twisted Poincar\'e algebra provides a simple example which illustrates the basic features of deformed symmetries and Fock space.  It was first introduced in \cite{Oeckl:2000eg, Chaichian:2004za} as the relevant algebraic structure to describe the relativistic symmetries of of a quantum field living on a specific type of non-commutative space-time, the Moyal plane
\begin{equation}
[x_{\mu},x_{\nu}]=i\theta_{\mu\nu}\, .
\end{equation} 
For this particular type of deformation the {\it modulation map} is given by what is technically called a ``Drinfeld twist" (for details see  \cite{Chaichian:2004za})
\begin{equation}
\mathcal{F}_{\theta}=\exp(-\frac{i}{2}\theta^{\alpha\beta}P_{\alpha}\otimes P_{\beta})\,. 
\end{equation} 
Such map obviously {\it does not} change the coproduct of translation generators since it commutes with them
\begin{equation}
\Delta_{\theta} P_{\mu} = \mathcal{F}_{\theta}\Delta P_{\mu}\mathcal{F}^{-1}_{\theta} \equiv\Delta P_{\mu}
\end{equation}
but it {\it does} change the coproduct of the generators of Lorentz rotations 
$\Delta_{\theta}M_{\mu\nu}\neq \Delta M_{\mu\nu}$, again for details we refer the reader to the review \cite{Akofor:2008ae}.  The important point we would like to stress is that even such mild deformation has an effect on the definition of multiparticle states and the construction of the Fock space of the theory.  Indeed now given two plane wave states $e_{\bf k_1}, e_{\bf k_2}\in \mathcal{H}_{\theta}$ the ordinary ``flip map" $\tau$ which exchanges the left and right component of elements of $\mathcal{H}\otimes\mathcal{H}$ {\it is no longer an intertwiner of representations of the twisted algbera} \cite{Akofor:2008ae}.  Instead we must replace $\tau$ by a twisted flip 
\begin{equation}
\tau_{\theta}(e_{\bf k_1}\otimes e_{\bf k_2})=\exp(-i\theta^{\alpha\beta}k_{1\alpha}\otimes k_{2\beta})(e_{\bf k_2}\otimes e_{\bf k_1})\, .
\end{equation}
Now in order to construct a generic bosonic n-particle state we need a rule to symmetrize tensor product states.  For example for a two particle state we would normally take the superposition 
\begin{equation}
|e_{\bf k_1}e_{\bf k_2}\rangle = 1/\sqrt{2}\,(e_{\bf k_1}\otimes e_{\bf k_2}+e_{\bf k_2}\otimes e_{\bf k_1})\equiv\frac{(1+\tau)}{\sqrt{2}}e_{\bf k_1}\otimes e_{\bf k_2}\, ,
\end{equation}
however in the twisted symmetry case since $\tau$ is no longer an intertwiner we must use the {\it twisted flip}
\begin{equation}
\tau_{\theta}=\mathcal{F}^{-1}_{\theta}\tau\mathcal{F}_{\theta}
\end{equation}
and the new two-particle state will be given by
\begin{equation}
|e_{\bf k_1}e_{\bf k_2}\rangle_{\theta}=\frac{(1+\tau_{\theta})}{\sqrt{2}}e_{\bf k_1}\otimes e_{\bf k_2}\, .
\end{equation}
This construction easily extends to $n$-particle states.  The non-trivial structure of the twisted Fock space amounts to the appearance of phases depending on the modes of the states involved and on the deformation parameter $\theta$.  Introducing creation and annihilation operators, $c_{\bf k}$ $c^{\dagger}_{\bf k}$ one can see that such ``momentum dependent" statistics is indeed encoded in the {\it non-canonical} commutation relations
\begin{equation}
c^{\dagger}_{\bf k_1}c^{\dagger}_{\bf k_2}-\exp(i\theta^{\alpha\beta}k_{1\alpha}\otimes k_{2\beta})
c^{\dagger}_{\bf k_2}c^{\dagger}_{\bf k_1}=0
\end{equation}
\begin{equation}
c_{\bf k_1}c^{\dagger}_{\bf k_2}-\exp(-i\theta^{\alpha\beta}k_{1\alpha}\otimes k_{2\beta})
c^{\dagger}_{\bf k_2}c_{\bf k_1}=\delta_{\bf k_1 k_2}\, .
\end{equation}
After this flash review of twisted Fock space we turn in the next subsection to a ``stronger" form of deformation, the $\kappa$-deformation, and explore its possible physical consequences.

\subsection{$\kappa$-Fock space and hidden entanglement}
The first example of deformation of the algebra of relativistic symmetries, the $\kappa$-Poincar\'e algebra, was proposed more than fifteen years ago \cite{Lukierski:1992dt}.  Our main interest in such algebra is due to the fact that it shares an important feature with the deformed symmetry algebras that have been shown to emerge in the context of three-dimensional quantum gravity \cite{Bais:2002ye}: the coproduct for translation generators (unlike the $\theta$ case) is now deformed.  Indeed, in the specific variant of this deformation which is of interest to us, the coproduct is given by
\begin{equation}
\label{kcop}
\Delta_{\kappa}{\bf P}={\bf P}\otimes 1+e^{-P_0/\kappa}\otimes {\bf P}
\end{equation}
where the dimensionful deformation parameter $\kappa$ is set at a given UV scale, usually the Planck energy $E_p$.
Other salient features of the $\kappa$-Poincar\'e algebra are the deformed positive and negative mass-shells (for simplicity we write it for the massless case)
\begin{equation}
\label{mass}
\epsilon^{\pm}({\bf p})=-\kappa\log\left(1\mp\frac{|{\bf p}|}{\kappa}\right)\, ,
\end{equation}
and a deformed action of boosts on linear momenta 
\begin{equation}
[N_j, P_l]=i\delta_{lj}\Big( \frac{\kappa}{2}  \left(1-e^{-\frac{2 P_0}{\kappa}} \right) +\frac{1}{2 \kappa} {\bf P}^2 \Big)+  \frac{i}{\kappa}P_l P_j \, .
\end{equation}
Note how at low energies ($\kappa\rightarrow \infty$) the deformation switches off and we recover the usual Poincar\'e algebra and its representations (for more details on the $\kappa$-Poincar\'e algebra and related field theory see e.g. \cite{Arzano:2007qp} and references therein).  As noted in \cite{Arzano:2008bt} (see also \cite{Bu:2006dm, Govindarajan:2008qa}) the deformed coproduct $\Delta_{\kappa}{\bf P}$ can again be obtained from the undeformed one using a ``modulation map"
\begin{equation}
\label{ktwist}
\mathcal{F}_{\kappa}=\exp\left(\frac{1}{\kappa}P_0\otimes P_j\frac{\partial}{\partial P_j}\right)\, .
\end{equation}
However in this case (unlike for the twisted coproduct discussed previously) such map {\it is not} a Drinfeld twist.  This seems to raise certain mathematical inconsistencies when dealing with bosonic $n$-particle states for $n\geq 3$ \cite{Young:2008zm}.  We will ignore such issues at this level and consider the features of $\kappa$-deformations listed above as the building blocks for a basic $\kappa$-Fock space construction which will allow us to explore the new physics of $\kappa$-deformed free fields.\\
We start with the construction of the one-particle Hilbert space.  The most straightforward way to characterize such space in this context is to consider functions on the deformed mass-shell\footnote{In the undeformed case there is a natural map via Fourier transform between functions on the mass-shell and solutions of the Klein-Gordon equation which allows to define the Hilbert space in two alternative ways.} $M^{\kappa}$ given by (\ref{mass}) equipped with the inner product (see \cite{Arzano:2007gr}) 
\begin{equation}
\label{kinnp}
<\phi_1,\phi_2>_{\kappa}=-i\omega\circ\mathcal{F}_{\kappa}(\bar{\phi}_1\otimes\phi_2)
\end{equation}
where $\omega$ is the ordinary symplectic product on the {\it undeformed} mass shell.  The next step is to find a suitable subspace on which the inner product above is positive definite.  While in the undeformed case it is enough to restrict to the positive mass shell in the present case one notices that for ``transplanckian" ($|\vec{p}|>\kappa$) modes (\ref{kinnp}) is no longer positive definite.  In other words the $\kappa$-one-particle Hilbert space will be given by the positive energy mass-shell truncated at $\kappa$, $M^{\kappa +}_{|{\bf k}|\leq\kappa} \subset M^{\kappa}$ equipped with the deformed inner product (\ref{kinnp}).\\  Turning to the multiparticle sector of the theory we notice that, as in the twisted case, ordinary symmetrization can not be used to construct $\kappa$-multiparticle states.  Indeed it is easily seen that, for example, if we consider the standard two-particle state
\begin{equation}
\label{symmst}
1/\sqrt{2}\,(|{\bf k_1}\rangle\otimes |{\bf k_2}\rangle+|{\bf k_2}\rangle\otimes |{\bf k_1}\rangle)
\end{equation}
this {\it is not} an eigenstate of the spatial translation generators ${\bf P}$ due to the non-trivial coproduct (\ref{kcop}).  Again in this case we have to resort to a non-trivial ``momentum dependent symmetrization".  One possibility is to proceed in analogy with the twisted case and make use of the ``modulation map" (\ref{ktwist}).  We will then construct a symmetrized state using the ``modulated flip" operator  $\tau^{\kappa}=\mathcal{F}_{\kappa}\tau\mathcal{F}^{-1}_{\kappa}$ whose action on a tensor product state is given by  
\begin{equation}
\tau^{\kappa}(|{\bf k_1}\rangle\otimes |{\bf k_2}\rangle)=|(1-\epsilon_1)\,{\bf k_2}\rangle\otimes |
(1-\epsilon_2)^{-1}\,{\bf k_1}\rangle\,,
\end{equation}
where $\epsilon_i=\frac{|{\bf k}_i|}{\kappa}$.  As an illustrative example consider the construction of a two-particle state.  Given a set of two one-particle states $|{\bf k_1}\rangle$ and $|{\bf k_2}\rangle$ we will now have {\it two distinct} two particle states
\begin{eqnarray}
|{\bf k_1k_2}\rangle_{\kappa}=&\frac{1}{\sqrt{2}}\left[ |\,{\bf k_1}\rangle\otimes\,|\,{\bf k_2}\rangle+|\,(1-\epsilon_1){\bf k_2}\rangle\otimes\,|\,(1-\epsilon_2)^{-1}{\bf k_1}\rangle\right]\\
|{\bf k_2k_1}\rangle_{\kappa}=&\frac{1}{\sqrt{2}}\left[ |\,{\bf k_2}\rangle\otimes\,|\,{\bf k_1}\rangle+|\,(1-\epsilon_2){\bf k_1}\rangle\otimes\,|\,(1-\epsilon_1)^{-1}{\bf k_2}\rangle\right]
\end{eqnarray}
with {\it same energy and different linear momentum} given respectively by 
\begin{eqnarray}
{\bf K_{12}}={\bf k_1}\dot{+}{\bf k_2}=&{\bf k_1}+(1-\epsilon_1){\bf k_2}\\
{\bf K_{21}}={\bf k_2}\dot{+}{\bf k_1}=&{\bf k_2}+(1-\epsilon_2){\bf k_1}\, .
\end{eqnarray}
In general given $n$-different modes one has $n!$ {\it different} $n$-particle states, one for each permutation of the $n$ modes  ${\bf k_1}\, ,{\bf k_2}\, ...\,{\bf k_n}$.  We see that the non-trivial algebraic structure of $\kappa$-translations endows the Fock space with a ``fine structure".  Indeed the different states (which in the absence of deformation would be degenerate) can in principle be distinguished by measuring their momentum splitting.  For example for the two particle states above the momentum splitting is given by
\begin{equation}
|\Delta{\bf K_{12}}|\equiv |{\bf K_{12}}-{\bf K_{21}}| =\frac{1}{\kappa}|{\bf k_{1}}|{\bf k_{2}}|-{\bf k_{2}}|{\bf k_{1}}||\le\frac{2}{\kappa}|{\bf k_{1}}||{\bf k_{2}}|
\end{equation}
which is of order $|{\bf k_{i}}|^2/\kappa$.  In other words the 2-particle Hilbert (sub)space becomes  
$\mathcal H^2_{\kappa} \cong \mathcal S_2\mathcal H^2 \otimes {\bf C}^2$, where $\mathcal S_2 \mathcal H^2$ is the ordinary symmetrized 2-particle Hilbert space.  We can thus write our states as
\begin{eqnarray}
|\epsilon\rangle\otimes|\uparrow\rangle &=& |{\bf k_{1}}{\bf k_{2}}\rangle _{\kappa}\\
|\epsilon\rangle\otimes|\downarrow\rangle &=&|{\bf k_{2}}{\bf k_{1}}\rangle _{\kappa}
\end{eqnarray}
with $\epsilon=\epsilon({\bf k_{1}})+\epsilon({\bf k_{2}})$.   Due to this additional structure we can now have states in which the macroscopic degrees of freedom of $S_2\mathcal H^2$ can be {\it entagled} with the ``planckian" degrees of freedom ${\bf C}^2$ as e.g. in the following state, superposition of two-particle states with total energies $\epsilon_A=\epsilon({\bf k_1}_A)+\epsilon({\bf k_2}_A)$ and $\epsilon_B=\epsilon({\bf k_1}_B)+\epsilon({\bf k_2}_B)$
\begin{equation}
|\Psi\rangle = 1/\sqrt{2}(|\epsilon_A\rangle\otimes|\uparrow\rangle
+|\epsilon_B\rangle\otimes|\downarrow\rangle)\, .
\end{equation}
The remarkable fact is that the possibility of this micro-macro entanglement renders possible interesting phenomena of decoherence \cite{Arzano:2008yc}.  To illustrate this consider the unitary evolution of a quantum system $\rho(t) = U(t) \rho(0) U^\dagger(t)$ whose states are represented by our $\kappa$-Fock space.  We start with a {\it pure state} $\rho(0)$ factorized with respect to the bipartition in $\mathcal H^{n}_\kappa\cong \mathcal S_n\mathcal H^n \otimes {\bf C}^{n}$.  If $U(t)$ acts as an ``entangling gate", the state $\rho(t)$ will be entangled.  A {\it macroscopic observer} who is not able to resolve the planckian degrees of freedom at the beginning will see the reduced system in a {\it pure state} $\rho_{obs}(0)=\mbox{Tr}_{Pl}\rho(0)$.  As the system evolves she will see the system in the \underline{mixed state}
\begin{equation} \rho_{obs}(t) = \mbox{Tr}_{Pl}\rho(t)=
\mbox{Tr}_{Pl}\left[ U(t) \rho(0) U^\dagger(t) \right].
\end{equation}
For the macroscopic observer, the evolution is not unitary!  This shows how  $\kappa$-Fock space provides a simple example of a quantum system which exhibits decoherence due to the presence of hidden ``planckian" degrees of freedom.  A similar scenario  has been advocated for certain models which could provide experimental signatures of quantum gravity (see Prof. Mavromatos contribution to these Proceedings).
\section{Conclusion}
We have provided a brief overview of the interplay between deformations of relativistic symmetries and field quantization.  In particular we discussed how the introduction of a deformation leads to non-trivial Fock space construction and a ``momentum dependent"  statistics.  In the specific case of $\kappa$-deformations, where the Leibnitz rule for spatial translation generators is deformed, we showed that the new structure of $\kappa$-Fock space allows for decoherence due to entanglement between ``macroscopic" and ``planckian" degrees of freedom.  It is important at this point to investigate whether this decoherence phenomena might lead to effects which could be checked in the real world.  Work is in progress in order to answer such question.

%%%%%%%%%%%%%%%%%%%%%%%%%%%%%%%%%%%%%%%%%%%%%%%%
%% BACKMATTER
%%%%%%%%%%%%%%%%%%%%%%%%%%%%%%%%%%%%%%%%%%%%%%%%

\begin{acknowledgments} 
I would like to thank the organizers of the XXV Max Born Symposium (Wroclaw, Poland,
29th June- 3rd July 2009) for the invitation and all the participants for providing such a stimulating environment.\\
Research at Perimeter Institute for Theoretical Physics is supported in part by the Government of Canada through NSERC and by the Province of Ontario through MRI.  
\end{acknowledgments}

\end{document}